\documentclass{PoS}

\title{Simulating linear covariant gauges on the lattice: a new approach}

\ShortTitle{Lattice linear covariant gauge}

\author{\speaker{Attilio Cucchieri}\\
        Instituto de F\'{\i}sica de S\~ao Carlos, Universidade de
        S\~ao Paulo \\ Caixa Postal 369, 13560-970 S\~ao Carlos, SP, Brazil \\
        E-mail: \email{attilio@ifsc.usp.br}}

\author{Tereza Mendes\\
        Instituto de F\'{\i}sica de S\~ao Carlos, Universidade de
        S\~ao Paulo \\ Caixa Postal 369, 13560-970 S\~ao Carlos, SP, Brazil \\
        E-mail: \email{mendes@ifsc.usp.br}}

\author{Elton M.\ da S.\ Santos\\
        Instituto de F\'{\i}sica de S\~ao Carlos, Universidade de
        S\~ao Paulo \\ Caixa Postal 369, 13560-970 S\~ao Carlos, SP, Brazil \\
        and \\
        Instituto de Educa\c c\~ao Agricultura e Ambiente, Campus Vale do Rio Madeira \\
        Universidade Federal do Amazonas, 69800-000 Humait\'a, AM, Brazil \\
        E-mail: \email{elton@ifsc.usp.br}}

\abstract{We discuss a new lattice implementation of the linear covariant gauge,
recently introduced in \cite{Cucchieri:2009kk}.
In particular, we present details of the numerical procedure for fixing the gauge.
We also report on preliminary results for the transverse and longitudinal gluon propagators
for the SU(2) gauge group in four space-time dimensions.}

\FullConference{International Workshop on QCD Green's Functions, Confinement,
                and Phenomenology - QCD-TNT09\\ September 07 - 11 2009\\
		ECT Trento, Italy}

\begin{document}


\section{Introduction}

The infrared behavior of Green's functions in Landau gauge has been
the topic of numerous lattice studies by several groups in the past few
years. Particular attention has been devoted to the gluon and ghost
propagators, whose infrared behavior is at the heart of the Gribov-Zwanziger
confinement scenario \cite{Gribov:1977wm,Zwanziger:1993dh,Zwanziger:2009je,
Cucchieri:2008yp}. There is now a consistent picture --- from extensive
numerical simulations on very large lattices \cite{Bogolubsky:2007ud,
Cucchieri:2007md,Sternbeck:2007ug,Cucchieri:2007rg, Cucchieri:2008fc,
Bornyakov:2008yx,Bogolubsky:2009dc,Oliveira:2007dy,Gong:2008td} ---
that (in three and in four space-time dimensions) the Landau gluon propagator
shows a massive solution at small momenta and that the Landau ghost propagator
is essentially free in the same limit. These results are not in agreement
with the original Gribov-Zwanziger scenario \cite{Gribov:1977wm,
Zwanziger:1993dh} but they can be explained in the so-called {\em refined
Gribov-Zwanziger framework} \cite{Dudal:2009bf}. Let us also recall that
a massive gluon allows a better description of experimental data \cite{Natale:2009uz}
and it has been related to color confinement by various authors
\cite{Cornwall:1979hz,Chaichian:2006bn}.

Since the evaluation of Green's functions depends on the gauge condition,
it is important to consider different gauges in order to obtain a clear
(possibly gauge-independent) picture of color confinement. Needless to say,
this investigation should be carried out at the nonperturbative level. This is done
from first principles using lattice simulations. In addition to the Landau gauge
case cited above, numerical studies of Green's functions have also been done in Coulomb
gauge \cite{Cucchieri:2006hi,Burgio:2008jr,Greensite:2009eb}, $\lambda$-gauge (a
gauge that interpolates between Landau and Coulomb) \cite{Cucchieri:2007uj} and
maximally Abelian gauge \cite{Bornyakov:2003ee, Mendes:2006kc}. For interesting
recent comparisons of results in Landau and in Coulomb gauge see \cite{Greensite:2009iv,
Burgio:2009xp}. 

On the other hand, the linear covariant gauge --- which is a
generalization of Landau gauge --- proved for a long time quite hostile
to the lattice approach \cite{Giusti:1996kf,Giusti:1999wz,Giusti:1999im,
Giusti:1999cw,Giusti:2000yc,Giusti:2001kr,Cucchieri:2008zx,Mendes:2008ux}.
Recently we have introduced a new implementation of the linear covariant
gauge on the lattice \cite{Cucchieri:2009kk}, based on a minimizing functional
that extends in a natural way the Landau case while preserving all the
properties of the continuum formulation. Let us note that, having a minimizing functional
for the linear covariant gauge allows a numerical investigation of the first
Gribov region $\Omega$ for the case of gauge parameter $\xi \neq 0$. Such an
investigation has been done analytically in \cite{Sobreiro:2005vn}, for a small
value of $\xi$, but a similar numerical study is still lacking.
At the same, a numerical investigation of the infrared behavior of gluon and ghost
propagators at $\xi \neq 0$ could provide important inputs for analytic studies
based on Dyson-Schwinger equations \cite{Alkofer:2003jr,Aguilar:2007nf}. Finally,
it has been recently proven \cite{Binosi:2002ft,Binosi:2003rr,Binosi:2009qm}
that the background-field Feynman gauge is equivalent
(to all orders) to the pinch technique \cite{Binosi:2009qm,Cornwall:1981zr}.
Thus, numerical studies using the Feynman gauge, which corresponds to the value $\xi = 1$,
will allow a nonperturbative evaluation of the gauge-invariant off-shell Green's
functions of the pinch technique \cite{Cornwall:2009as}.


\section{Linear Covariant Gauge}
\label{sec:lcg}

In the linear covariant gauge the gluon field $A_{\mu}^b(x)$
satisfies (in the continuum) the relation
\begin{equation} 
   \partial_{\mu} A_{\mu}^b(x) \;=\; \Lambda^b(x)
\label{eq:Feynmancontinuo}
\;\mbox{,}
\end{equation}
where $\Lambda^b(x)$ are real-valued functions generated using a Gaussian
distribution
\begin{equation}
P\left[\Lambda^b(x)\right] \;\sim\; exp{\left\{ - \frac{1}{2 \xi}\, \sum_b\,
\left[ \Lambda^b(x) \right]^2 \right\} }
\label{eq:gaussian}
\end{equation}
with width $\sqrt{\xi}$.

The limit $\xi \to 0$ corresponds to the standard Landau gauge.
In this case, the gauge condition is (classically) equivalent
to the Lorenz-gauge (sometimes mistakenly called Lorentz-gauge)
condition \cite{Jackson:2001ia}
\begin{equation} 
   \partial_{\mu} A_{\mu}^b(x) \;=\; 0 \; . 
\label{eq:Landaucontinuo}
\end{equation}
This condition can be imposed by minimizing the functional
\begin{equation}
  {\cal E}_{LG}\{A^{g}\} \; \propto \;
     \int \, d^4x \; \sum_{\mu, b} \left[ (A^{g})_{\mu}^b(x) \right]^2
\label{eq:Landau}
\end{equation}
with respect to the gauge transformations $\{g(x)\}$. Let us recall here
that, from the second variation of ${\cal E}_{LG}\{A^{g}\}$, we can define
the Faddeev-Popov operator ${\cal M}$. Then, for the gauge-fixed configurations,
i.e.\ for local minima of ${\cal E}_{LG}\{A^{g}\}$, we have that this operator
is positive-definite. This set of local minima defines the
first Gribov region $\Omega$ \cite{Gribov:1977wm,Zwanziger:1993dh}.

In Ref.\ \cite{Giusti:1996kf} it was shown that a similar minimizing
functional ${\cal E}_{LCG}\{A^{g}\}$ for the linear covariant gauge
--- i.e.\ for $\xi \neq 0$ --- does not exist. Indeed, if it existed,
we could write
\begin{equation}
  {\cal E}_{LCG}[A^{g},\Lambda] \; =  \; {\cal E}_{LG}[A^{g}]
          \,+\, {\cal F}[A^{g},\Lambda] \; ,
\end{equation}
for some functional ${\cal F}[A^{g},\Lambda]$. Then, the second
variation of ${\cal E}_{LCG}$ with respect to the gauge transformation
$g(x) = e^{iw(x)}$ would satisfy the relation
\begin{equation}
 \frac{\partial^2 {\cal E}_{LCG}}{\partial w^b(x) \partial w^c(y)} \; = \; 
 \frac{\partial^2 {\cal E}_{LCG}}{\partial w^c(y) \partial w^b(x)} \; .
\end{equation}
On the other hand, one can show that these two terms are, respectively,
proportional to the structure functions $f^{acb}$ and $f^{abc}$. Since these functions
are completely anti-symmetric in the color indices, this equality
cannot be realized \cite{Giusti:1996kf}.


\section{Bypassing the No-Go Condition}

One can of course avoid the no-go condition above by relaxing its hypotheses. For
example, one can consider a different gauge condition, such as
\begin{equation}
 F [ \, \partial_{\mu} A_{\mu}^b(x) \,-\, \Lambda^b(x) \, ] \; = \; 0
\label{eq:fnew}
\end{equation}
with $F[0]=0$, for which a minimizing functional exists. Indeed, it
has been shown in Ref.\ \cite{Giusti:1996kf} that the minimizing functional
\begin{equation}
      \int \, d^4x \;\sum_{\mu, b} \left\{
   \left[ \partial_{\mu} A_{\mu}^b(x) - \Lambda^b(x) \right]^2
             \right\}
\label{eq:giusti}
\end{equation}
allows one to impose the gauge condition
\begin{equation}
D_{\nu}^{ab} \partial_{\nu} \left[ \partial_{\mu} A_{\mu}^b(x)
        - \Lambda^b(x) \right] \, = \, 0 \; ,
\end{equation}
where $\, D_{\nu}^{ab} \,$ is the covariant derivative.

On the other hand, the use of a different gauge condition introduces new problems \cite{Giusti:1996kf}.
For example, one can bring in spurious solutions, corresponding to $F[s]=0$
for $s \neq 0$. In the above case, these solutions are the zeros of the operator
$\, D_{\nu}^{ab} \partial_{\nu} \, $. Also, in general, the second variation
of the minimizing functional, or equivalently the first variation of the
gauge-fixing condition (\ref{eq:fnew}), does not correspond to the Faddeev-Popov
operator ${\cal M}\,=\,- \partial_{\mu} D_{\mu}^{ab}$ of the usual linear covariant
gauge. Finally, the lattice discretization of the functional (\ref{eq:giusti}) is
not linear in the gauge transformation $\{ g(x) \}$. This makes the numerical
minimization difficult and one has to rely on a specific discretization of the
minimizing functional \cite{Giusti:1999wz,Giusti:1999im} in order to make the lattice
approach feasible.

More recently \cite{Cucchieri:2008zx} the no-go condition has been overcome
by avoiding the use of a minimizing functional ${\cal E}_{LCG}\{A^{g}\}$.
To this end, following the perturbative definition of the linear covariant gauge
in the continuum, one first fixes the gluon field to Landau gauge
$\partial_{\mu} A_{\mu}^b(x) = 0$. Then, one considers the equation 
\begin{equation}
\left( \partial_{\mu} D_{\mu}^{bc} \phi^c\right)(x) \, = \, \Lambda^b(x) \; .
\end{equation}
The solution $\phi^c(x)$ of this equation can be used as a generator of a second
gauge transformation. Note that, after fixing the lattice (or minimal) Landau gauge,
the operator $ - \partial_{\mu} D_{\mu}^{bc} $ is positive-definite and can be
easily inverted. For small $\phi^c(x)$, the final (gauge-transformed) gluon
field ${A'}_{\mu}^{b}(x)$ satisfies the condition
\begin{equation}
\partial_{\mu} {A'}_{\mu}^{b}(x) \, = \, \partial_{\mu}
    \left(A_\mu^b + D_\mu^{bc}\phi^c\right)(x) \, = \, \Lambda^b(x) \; .
\end{equation}
Of course, the above result is correct only for infinitesimal gauge
transformations. On the other hand, usually $\phi^c(x)$ is not small
in a numerical simulation. Indeed, numerical tests \cite{Cucchieri:2008zx}
have shown that the distributions of $\partial_{\mu} {A'}_\mu^{b}(x)$ 
and of $\Lambda^b(x)$ do not agree very well.  At the same time, the relation
$p^2 D_l(p^2) = \xi$, valid in the linear covariant gauge for the
longitudinal gluon propagator $D_l(p^2)$, is also not well verified by the data
at small momenta \cite{Cucchieri:2008zx}.


\section{A New Approach}
\label{sec:new}

Our new approach \cite{Cucchieri:2009kk} is based on removing an {\em implicit}
hypothesis of the no-go condition, i.e.\ that the gauge transformation
$\{ g(x) \}$ appears in the minimizing functional in the ``canonical'' way
$A^{g}$. Thus, we may look for a minimizing functional of
the type ${\cal E}_{LCG}\{A^{g}, g\}$ instead of simply ${\cal E}_{LCG}\{A^{g}\}$.
Indeed, the lattice linear covariant gauge condition can be obtained by
minimizing the functional\footnote{Note that solving a system of
equations $B \psi = \zeta$ is equivalent to minimizing the quadratic form
$ \frac{1}{2} \psi B \psi \, - \, \psi \, \zeta $.}
\begin{equation}
 {\cal E}_{LCG}\{U^{g}, g\} \; = \; {\cal E}_{LG}\{U^{g}\}
                   \, + \, \Re \; Tr \sum_x
                     \,  i\, g(x) \, \Lambda(x) \; ,
\label{eq:EFeynman}
\end{equation}
where
\begin{eqnarray}
\!\!\!\!\!\!\!\!\!{\cal E}_{LG}\{U^{g}\} \!\! & = & \! \!
              - \; \Re \; Tr \sum_{x, \mu}
   \; g(x) \, U_{\mu}(x) \, g^{\dagger}(x+ e_{\mu})
          \label{eq:ELandau2} 
\end{eqnarray}
is the minimizing functional for the lattice Landau gauge.
Here, the link variables $U_{\mu}(x)$ and the
site variable $g(x)$ are matrices belonging to the SU($N_{c}$) group
(in the fundamental representation). We also indicate with $\Re\,$ the
real part of a complex number and with $Tr$ the trace in color space.
Note that the functional ${\cal E}_{LCG}\{U^{g}, g\}$ is linear in the
gauge transformation $\{ g(x) \}$.

By considering a one-parameter
subgroup $ g(x,\tau) \, = \, \exp \left[ i \tau \gamma^{b}(x) \lambda^{b} \right] $
of the gauge transformation $\{ g(x) \}$ it is
easy to check that the stationarity condition
\begin{equation}
 \left. \frac{\partial {\cal E}_{LCG}}{\partial \tau} \right|_{\tau =0} = 0
 \qquad \qquad \forall \;\;\; \gamma^{b}(x)
\end{equation}
implies the lattice linear covariant gauge condition
\begin{equation}
 \nabla \cdot A^{b}(x) \, = \, \sum_{\mu} \, A_{\mu}^{b}(x) 
                        \, - \, A_{\mu}^{b}(x-e_{\mu}) \, = \,
   \Lambda^{b}(x)\; .
   \label{eq:Feynman}
\end{equation}
Here, $\, \lambda^{b} $ are the traceless Hermitian generators of the Lie algebra
of the SU($N_{c}$) gauge group, satisfying the usual normalization condition
\begin{equation}
 Tr \left( \lambda^{b} \lambda^{c} \right) \, = \, 2 \, \delta^{bc} \; .
\end{equation}
Also, we used $ \Lambda^{b}(x) = Tr \, [ \Lambda(x) \, \lambda^{b} ] \, $
and, similarly, $ \, A_{\mu}^{b}(x) = Tr \, [ A_{\mu}(x) \, \lambda^{b} ] \, $.

At the same time, the second variation (with respect to the parameter $\tau$)
of the term $\,i\, g(x) \, \Lambda(x)\,$, on the r.h.s.\ of Eq.\ (\ref{eq:EFeynman}),
is purely imaginary. Thus, it does not contribute to the second variation
of the functional ${\cal E}_{LCG}\{U^{g}, g\}$. This implies that, using the
above minimizing functional, one finds for the Faddeev-Popov matrix
$\, {\cal M} \,$ a discretized version of the usual Faddeev-Popov operator
$- \partial \cdot D$.


\begin{figure}[t]
\begin{center}
\vskip -1.8cm
\hskip -5mm
\includegraphics[scale=0.40]{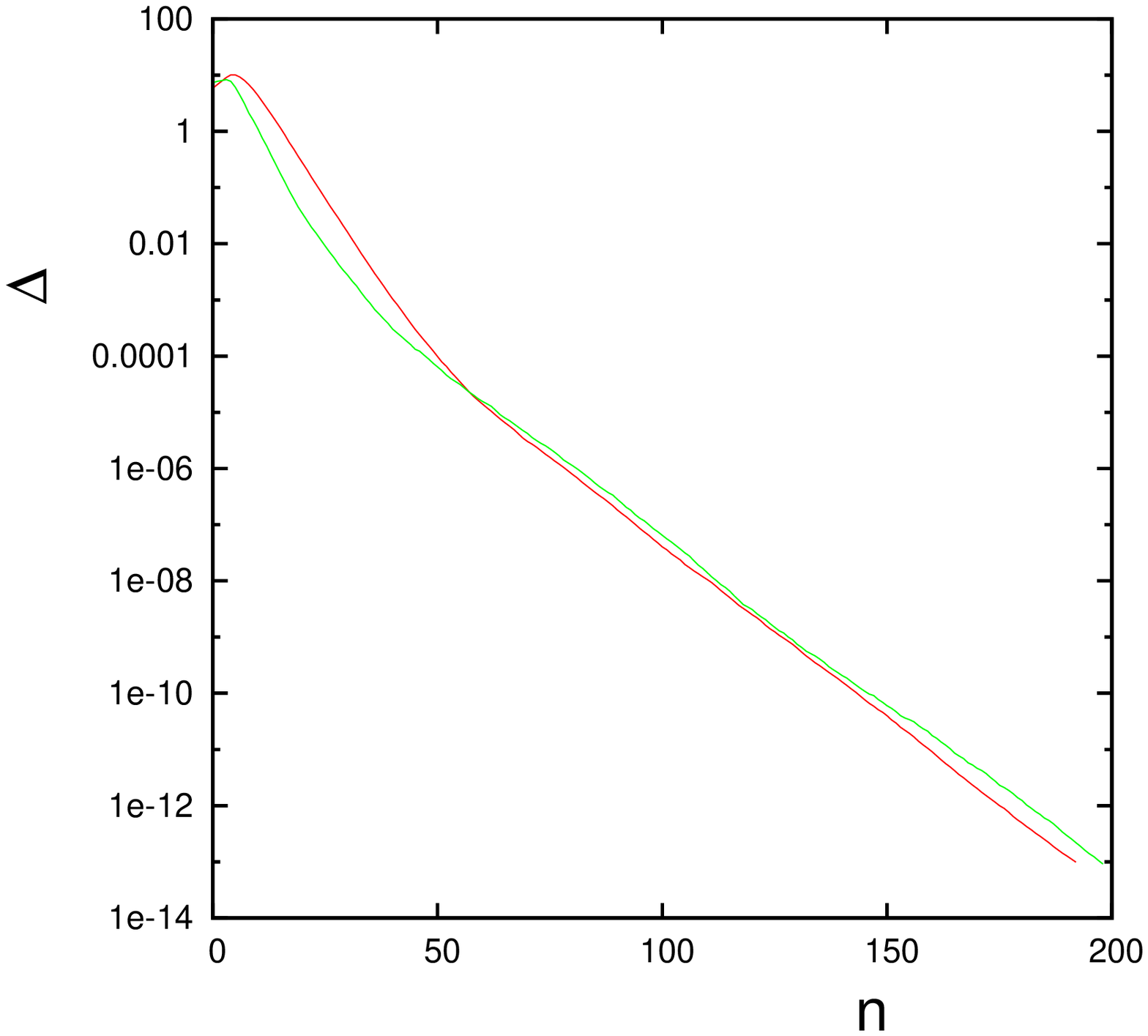}
\hskip 6mm
\includegraphics[scale=0.40]{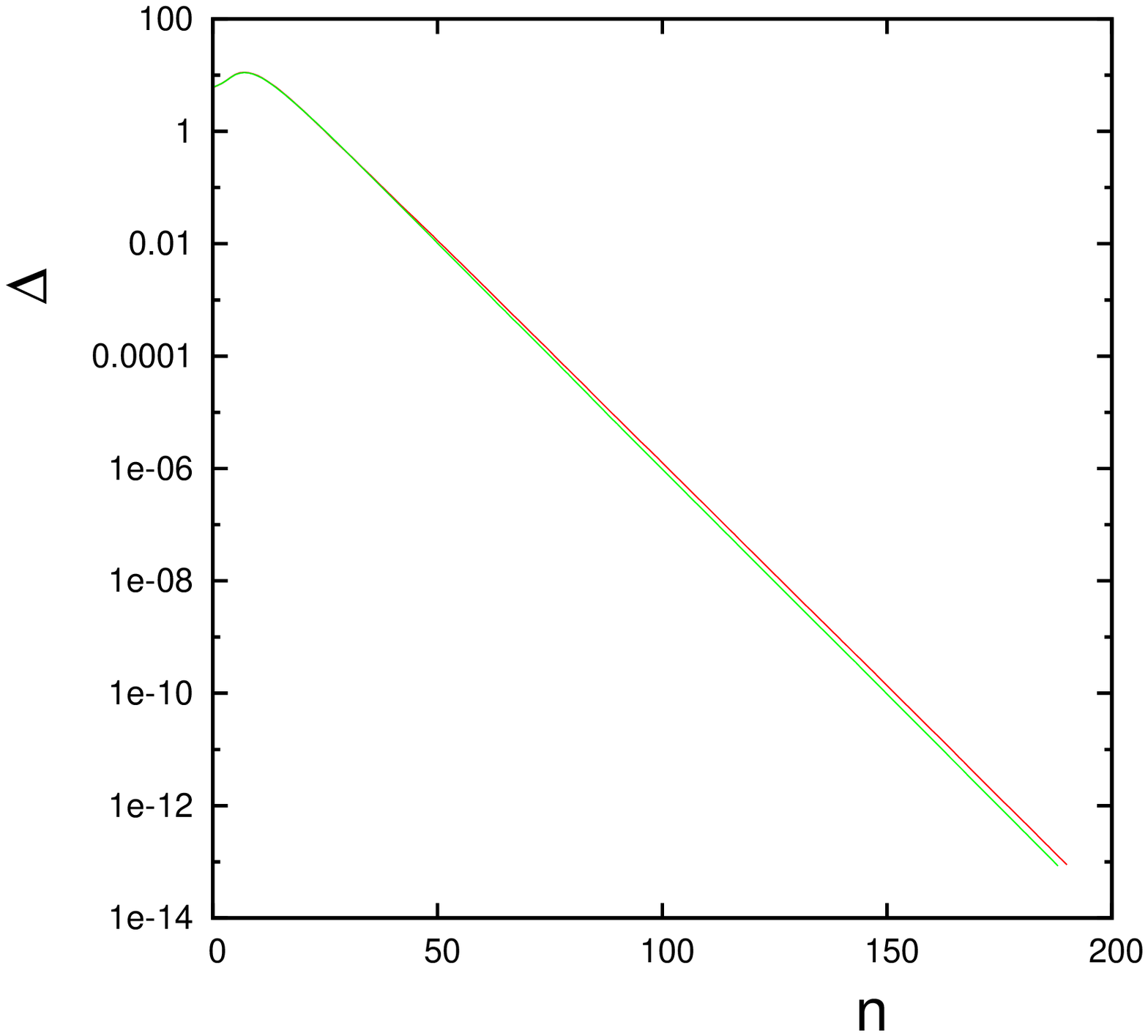}
\vskip -1.9cm
\caption{\label{fig:gfix}
  Convergence of the numerical gauge fixing. Here we report the value of
  $ \Delta = \sum_{x,b} [ \nabla \cdot A^{b}(x) - \Lambda^{b}(x) ]^2 $ as
  a function of the number of iterations $n$ for a given configuration.
  Left: $\beta = 4$, $V = 8^4$, $\xi = 0$ (red line) and $\xi = 0.5$
  (green line). Right: $\beta = 4$, $V = 16^4$, $\xi = 0$ (red line) and $\xi = 0.05$
  (green line). Note the logarithmic scale on the $y$ axis.
  }
\end{center}
\end{figure}

\section{Numerical Tests}

We have performed some numerical tests\footnote{Note that Eq.\ (\ref{eq:Feynman})
implies $ \sum_x \Lambda^{b}(x) = 0$. Thus, after generating the functions
$ \Lambda^{b}(x) $ using the Gaussian distribution (\ref{eq:gaussian}), one has
to remove possible zero modes from them.} with the functional (\ref{eq:EFeynman}),
using the sto\-chas\-tic-overrelaxation algorithm \cite{Cucchieri:1995pn,
Cucchieri:1996jm,Cucchieri:2003fb}.

For these first tests we considered the 4d SU(2) case at $\beta = 4$, for $V = 8^4$
and $16^4$, with $\xi = 0.01, 0.05, 0.1$ and $0.5$. The numerical gauge fixing works very
well when $\xi \neq 0$, at least for relatively small lattice volumes $V$ and
gauge parameter $\xi$ (see the next section). In Figure \ref{fig:gfix} (left panel)
we compare the gauge fixing for a given configuration at $\beta = 4$ and $V = 8^4$ for
the Landau case $\xi = 0$ (red line) and for $\xi = 0.5$ (green line). The rate of convergence
is essentially the same in the two cases. However, note that the tuning of the
stochastic-overrelaxation algorithm is different. Indeed, we found that,
for the chosen configuration, the best value for the stochastic-overrelaxation parameter
$p$ was about 0.73 in the Landau case and about 0.5 when $\xi = 0.5$. A similar result
(see Figure \ref{fig:gfix}, right panel) is obtained for $\beta = 4$ with $V = 16^4$ for the Landau
case ($\xi = 0$, red line) and for $\xi = 0.05$ (green line). In this case the best value
for $p$ was about 0.82 in the Landau case and about 0.81 when $\xi = 0.05$.  Let us note
that the functional $ {\cal E}_{LCG}\{U^{g}, g\}$ can be interpreted as a spin-glass
Hamiltonian for the {\em spin} variables $g(x)$ with a random interaction given by
$U_{\mu}(x)$, in a random external magnetic field $ \Lambda(x) $. The presence
of this magnetic field does not modify the convergence matrix \cite{Saad} or, as a
consequence, the behavior of the algorithm.\footnote{Note that, in the case $\beta = \infty$
\cite{Cucchieri:2003fb}, the Landau case corresponds to solving the Laplace
equation while the linear covariant gauge corresponds to solving a Poisson
equation. As is well-known, these two equations show the same critical behavior when
solved using a relaxation method such as Gauss-Seidel \cite{Sokal}.}

\begin{figure}[t]
\begin{center}
\vskip -1.8cm
\hskip -5mm
\includegraphics[scale=0.40]{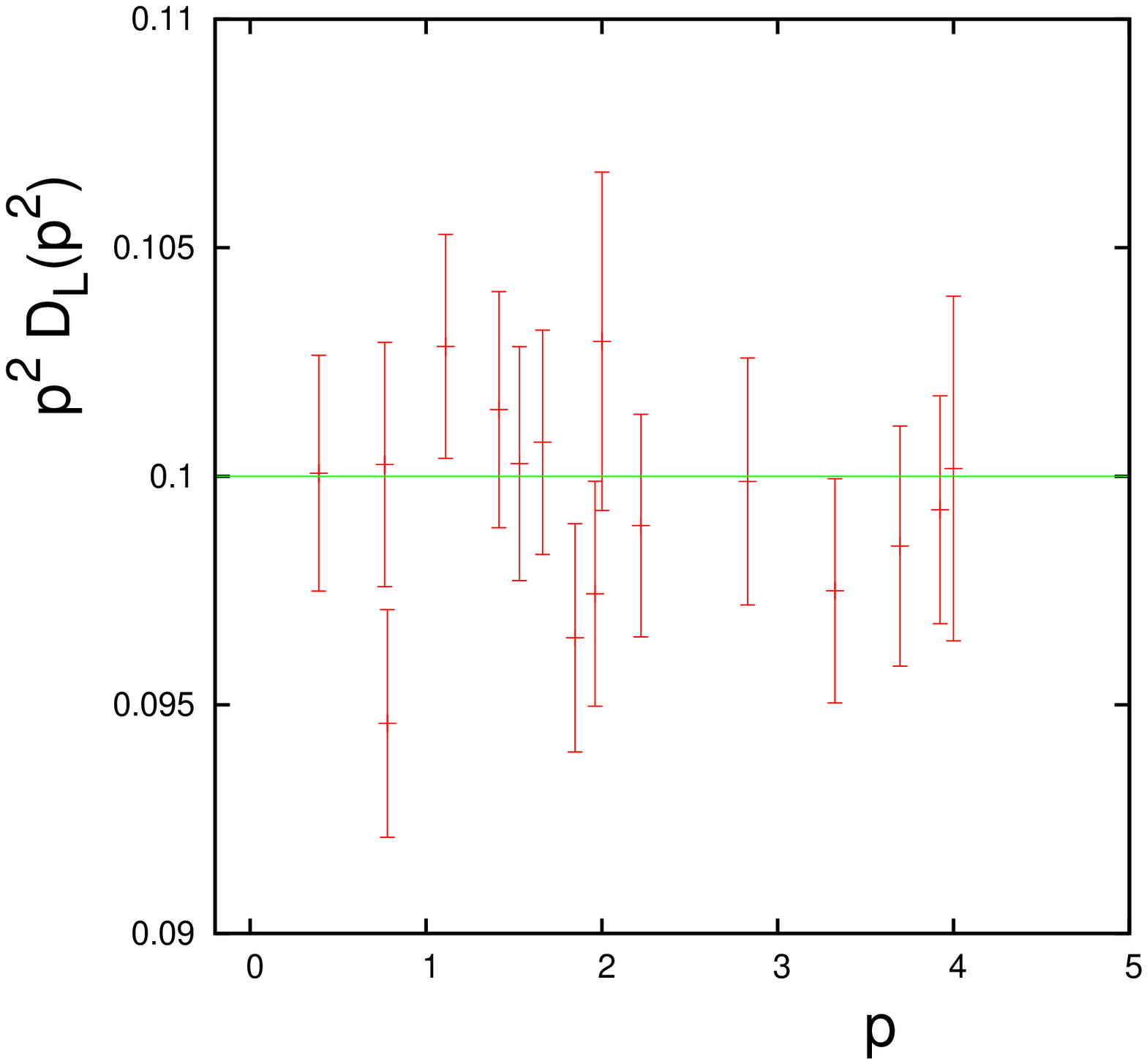}
\hskip 6mm
\includegraphics[scale=0.40]{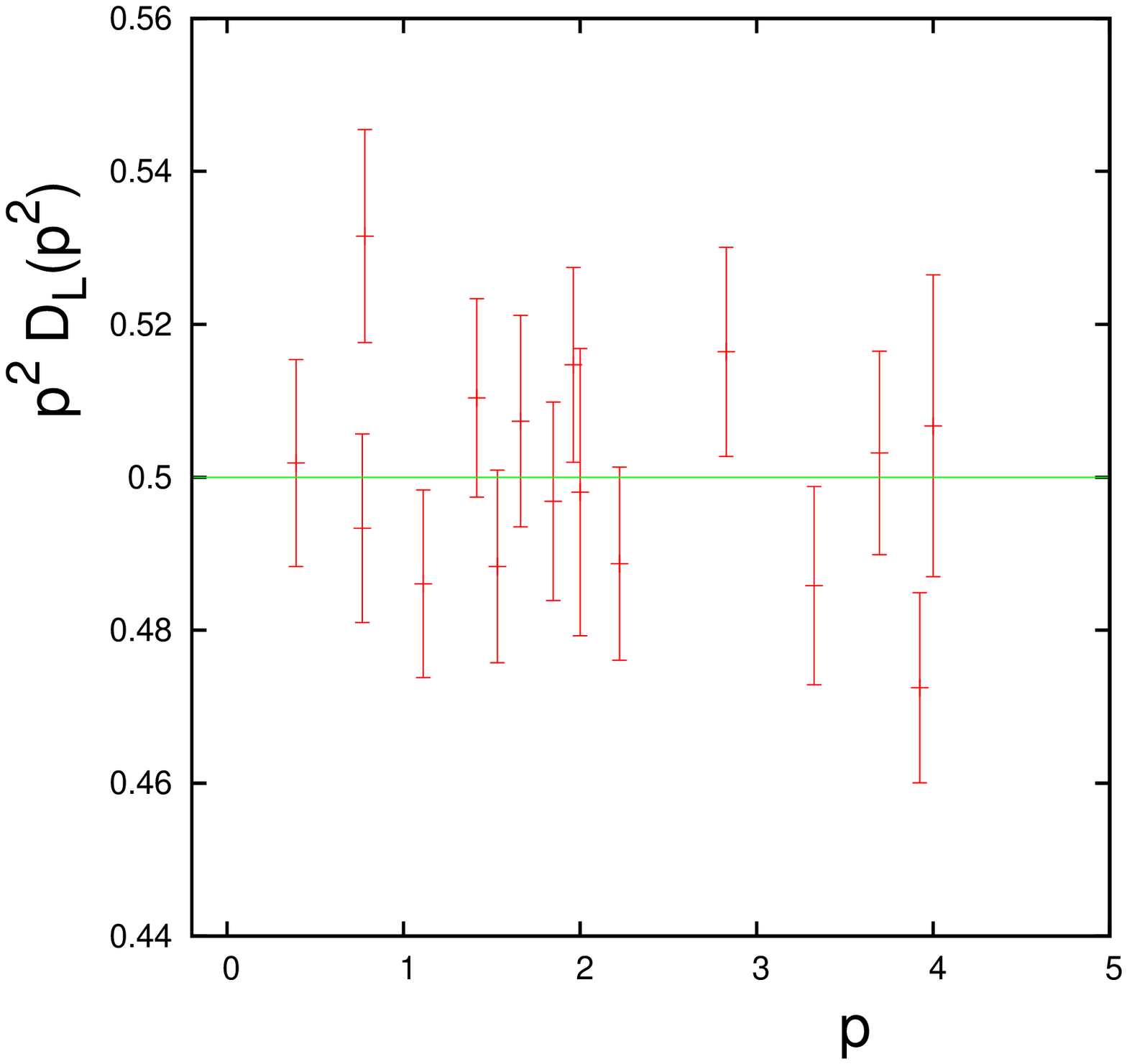}
\vskip -1.9cm
\caption{\label{fig:gfix3}
  The longitudinal gluon dressing function $p^2 D_l(p^2)$ as a function of the
  lattice momentum $p$ (in lattice units). We also show the predicted value
  $p^2 D_l(p^2) = \xi$.
  Left: $\beta = 4$, $V = 16^4$ and $\xi = 0.1$.
  Right: $\beta = 4$, $V = 16^4$ and $\xi = 0.5$.
  Note the relatively small range of values on the $y$ axis.
  }
\end{center}
\end{figure}

We also checked that the quantity $p^2 D_l(p^2)$ is constant within
statistical fluctuations in all cases considered. For $V = 16^4$
and $\xi = 0.1$ and $0.5$ the data are shown in Figure \ref{fig:gfix3}.
In these cases, a fit of the type $a/p^b$ for $D_l(p^2)$ gives
$a = 0.0994(7)$, $b = 2.003(9)$ with a $\chi^2/dof = 0.9$ when
$\xi = 0.1$ and $a = 0.502(5)$, $b = 2.01(1)$ with a $\chi^2/dof = 1.1$ when
$\xi = 0.5$. Similar fits have been obtained in the other cases.

\section{Discretization of the Gluon Field}
\label{sec:dis}

In the above tests we used the usual discretization
\begin{equation}
A_{\mu}(x) \, = \, \frac{\left[ U_{\mu}(x) - U_{\mu}^{\dagger}(x) \right]_{traceless}}{2 i}
\label{eq:discret}
\end{equation}
for the gluon field. However, one has to recall that, using this standard
(compact) discretization, the gluon field is bounded. On the other hand,
the functions $\Lambda^b(x)$ [see Eqs.\ (\ref{eq:Feynmancontinuo}) and
(\ref{eq:gaussian})] satisfy a Gaussian distribution, i.e.\ they are unbounded.
This can give rise to convergence problems \cite{Rank}. Moreover, the problem is
more severe for a larger width of the Gaussian distribution.

A possible way out of this problem is, of course, the use of different discretizations
of the gluon field $A_{\mu}(x)$, in order to improve the convergence of the minimizing
algorithms. To this end we also did some tests using the {\em angle} projection
\cite{Amemiya:1998jz} and the stereographic projection \cite{vonSmekal:2007ns}.
Note that, in the last case, the gluon field is in principle unbounded even for a finite
lattice spacing. We also stress that, in both cases, the numerical implementation
gets simplified if one uses the Cornell method \cite{Cucchieri:1995pn,
Cucchieri:1996jm,Cucchieri:2003fb} instead of the stochastic-overrelaxation algorithm,
as done in the simulations reported in the previous section.

We tested the standard discretization, the angle projection and the stereographic
projection using $V=8^4$, $\xi=0.01, 0.05,0.1,0.5,1.0$ and $\beta=2.2,2.3,\ldots,2.9,3.0$.
We found (see Table \ref{table-disc}) that the stereographic projection allows one
to simulate at slightly larger values of $\xi$, for a given lattice volume $V$ and
lattice coupling $\beta$, compared to the other two cases.\footnote{Note that, as explained
in the next section, the Gaussian distribution generated in the simulation actually
has squared width $\sigma = 4 \xi/\beta$ [in the SU(2) case].}

\begin{table}
\vskip -2mm
\caption{Smallest value of $\beta$ for which the numerical gauge-fixing algorithm
         showed convergence for the lattice volume $V=8^4$. Results are reported
         for the three different discretizations considered and for five different
         values of the gauge parameter $\xi$. For each case we used five different
         configurations.
\label{table-disc}}
\vskip 2mm
\begin{center}
\begin{tabular}{cccc}
\hline
 $\xi$ & stand.\ disc.\ & angle proj.\ & stereog.\ proj.\ \\
 \hline
 $0.01$ & 2.2 & 2.2 & 2.2 \\
 $0.05$ & 2.2 & 2.2 & 2.2 \\
 $0.1$  & 2.2 & 2.2 & 2.2 \\
 $0.5$  & 2.8 & 2.6 & 2.5 \\
 $1.0$  & --- & 3.0 & 2.5 \\
 \hline
\end{tabular}
\end{center}
\vskip 1mm
\end{table}


\section{Continuum Limit}
\label{sec:cont}

In Ref.\ \cite{Cucchieri:2008zx} it was shown that in the SU($N_c$) case, in order to
obtain the correct continuum limit, the functions $\Lambda^b(x)$ should be generated 
using a Gaussian distribution with width $\sqrt{\sigma} = \sqrt{2 N_c \xi/\beta}$,
instead of the width $\sqrt{\xi}$. Thus, for $\beta < 2 N_c$ the lattice width
$\sqrt{\sigma}$ is even larger than the continuum width $\sqrt{\xi}$. On the other hand,
one can always obtain a sufficiently small value for $\sigma$ by considering large
enough values of the lattice coupling $\beta$. However, if $\beta$ is too large, the
physical volume is too small (for a given lattice size) and one cannot study the
infrared limit of the theory.

Note that, in the SU(2) case, one has $\sigma = \xi$ only for $\beta = 4$, which corresponds
to a very small lattice spacing, i.e.\ $a \approx 0.001$ fm \cite{Bloch:2003sk}.
On the contrary, in the SU3 case, one has $\sigma = \xi$ for $\beta = 6$, corresponding
to a lattice spacing $a = 0.102$ fm \cite{Cucchieri:2007zm}, usually employed in lattice
numerical studies. Thus, simulations for the linear covariant gauge are probably easier
in the SU(3) case.

One should also recall that the gluon field $A$ and the gauge parameter
$\xi$ are (multiplicatively) renormalized by the same factor $Z_3$, i.e.\
$A_B = Z_3^{1/2} A_R$ and $\xi_B = Z_3 \, \xi_R$. Here, $B$ and $R$ indicate
bare and renormalized quantities respectively. This implies that, on the
lattice, data obtained for two different values of $\beta$ in
the scaling region --- e.g.\ $\beta_1$ and $\beta_2$ --- would give the
same (renormalized) propagator only if the multiplicative factor $R_Z =
Z_3(\beta_1) / Z_3(\beta_2)$ relating the propagators\footnote{See Ref.\
\cite{Leinweber:1998uu} for the case of Landau gauge.} also relates
the gauge parameters $\xi_1$ and $\xi_2$. Since the value of $R_Z$ is not
known a priori, one has to find numerically pairs of parameters $(\beta, \xi)$
yielding the same continuum renormalized propagators.


\section{Transverse Gluon Propagator}
\label{sec:results}

\begin{figure}[t]
\begin{center}
\vskip -1.8cm
\hskip -5mm
\includegraphics[scale=0.40]{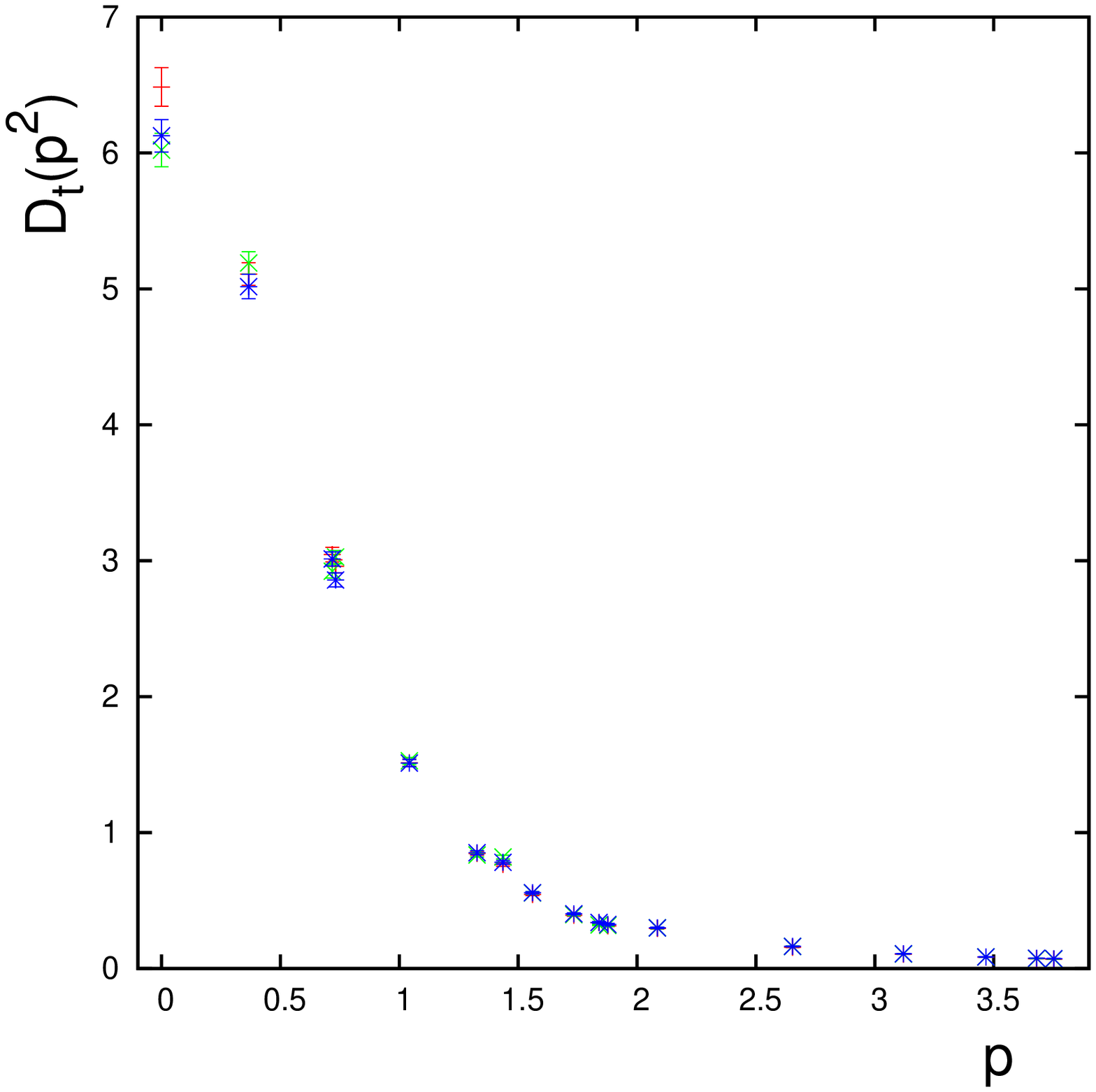}
\hskip 6mm
\includegraphics[scale=0.40]{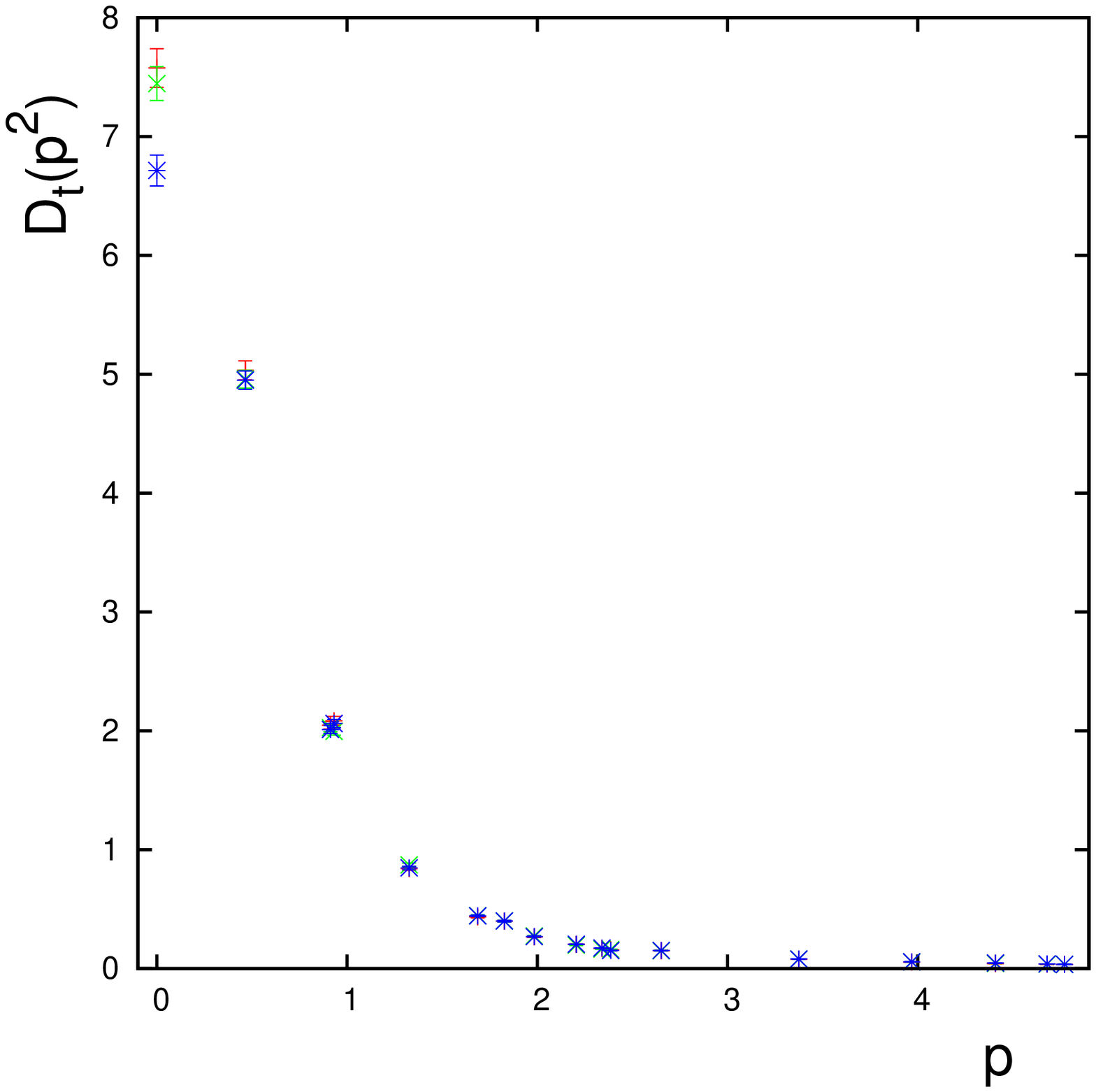}
\vskip -1.7cm
\caption{\label{fig:Dt16}
    Transverse gluon propagator $D_t(p^2)$ as a function of the
    momentum $p$ (both in physical units) for the lattice volume
    $V = 16^4$, $\beta = 2.2$ (left) and $\beta = 2.3$ (right), with
    $\xi = 0 \, (+, red), 0.05 \, (\times, green)$ and $0.1 \, (*, blue)$. 
  }
\end{center}
\end{figure}

\begin{figure}[t]
\begin{center}
\vskip -1.8cm
\hskip -5mm
\includegraphics[scale=0.40]{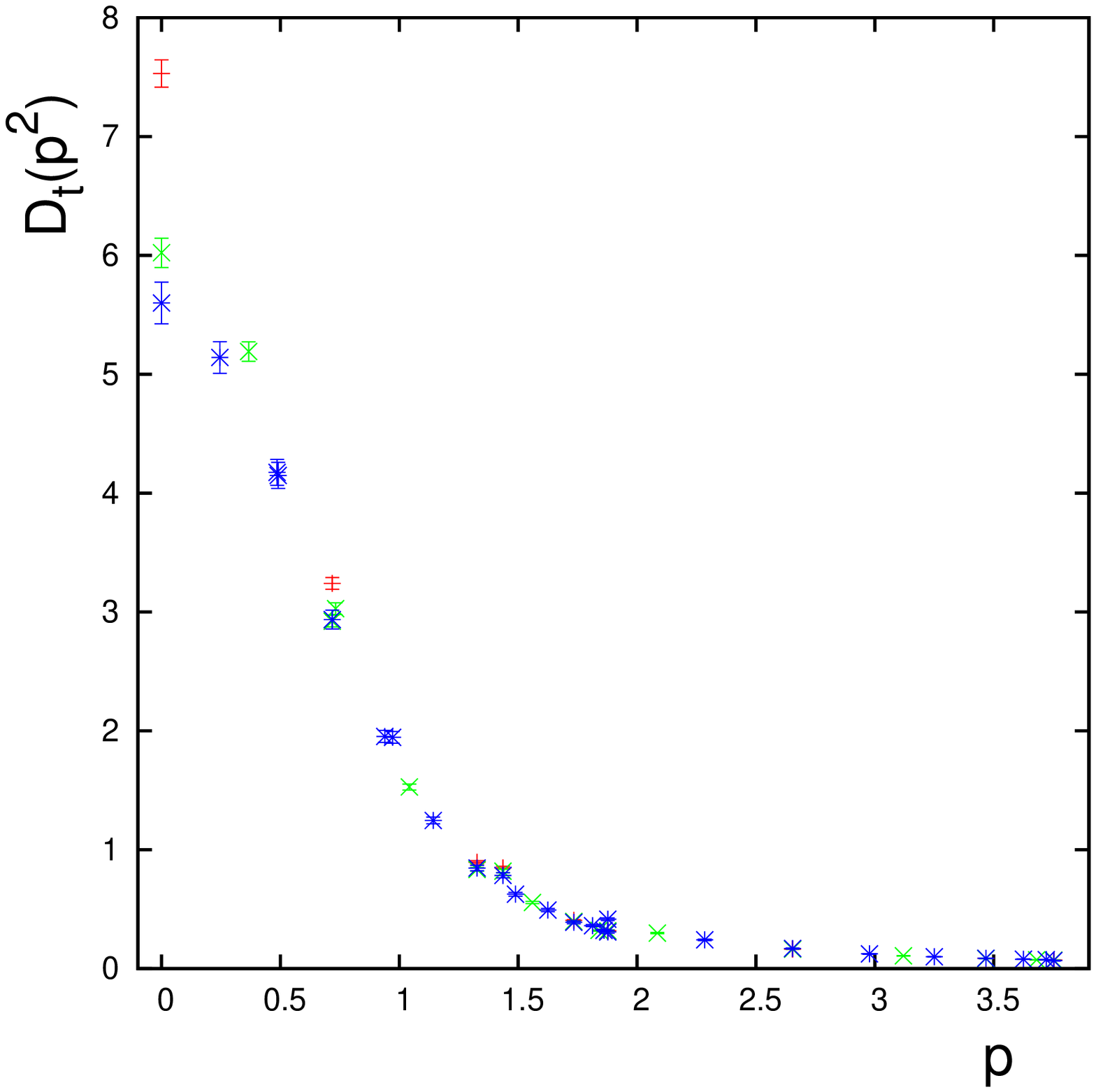}
\hskip 6mm
\includegraphics[scale=0.40]{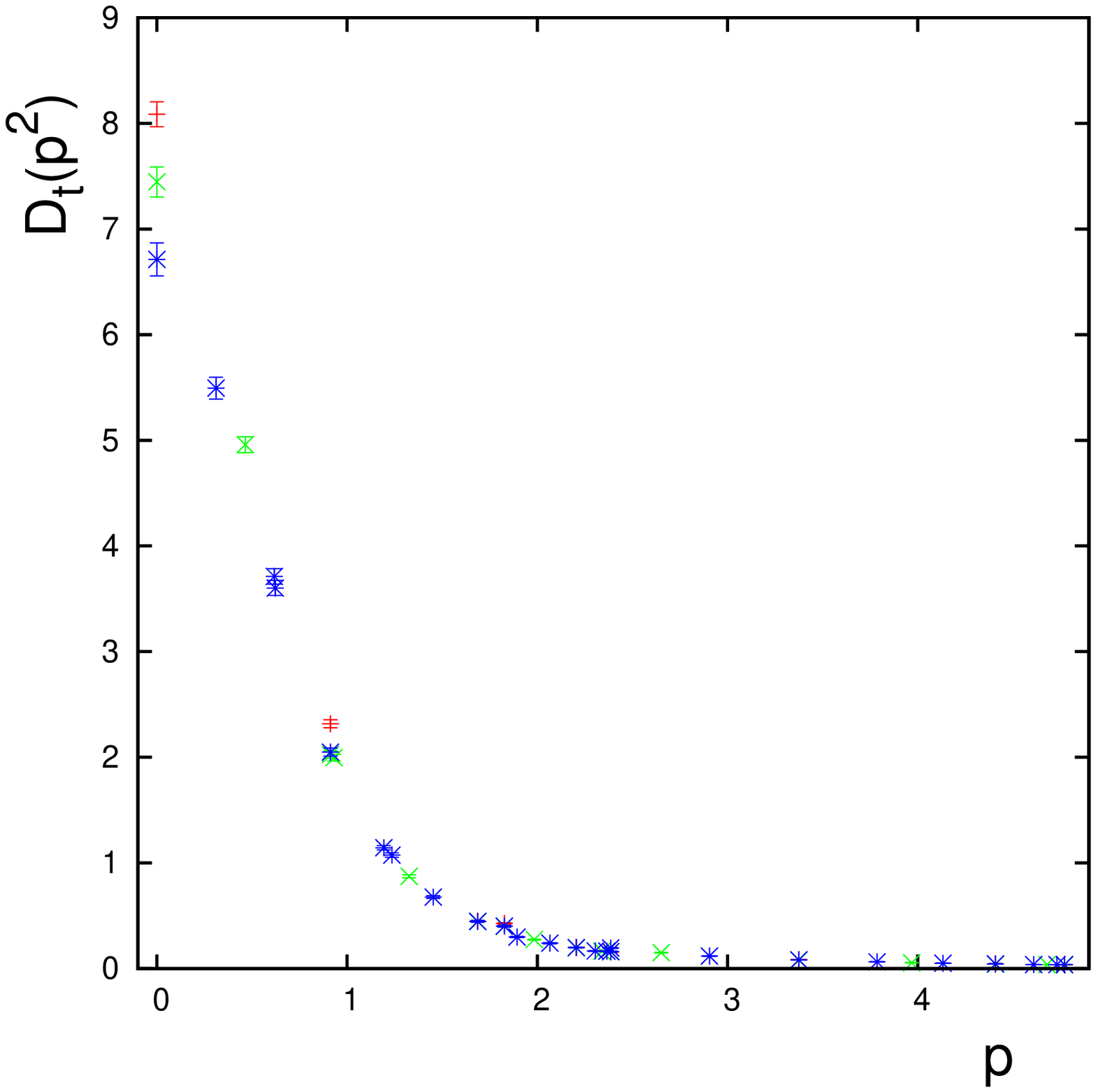}
\vskip -1.7cm
\caption{\label{fig:Dt}
    Transverse gluon propagator $D_t(p^2)$ as a function of the
    momentum $p$ (both in physical units) for the gauge coupling $\xi = 0.05$,
    $\beta = 2.2$ (left) and $\beta = 2.3$ (right), with the lattice volumes
    $V = 8^4 \, (+, red), 16^4 \, (\times, green)$ and $24^4 \, (*, blue)$.
  }
\end{center}
\end{figure}

In this section we present preliminary results for the momentum-space transverse
gluon propagator $D_t(p^2)$ for different values of $\xi$. Using the stereographic
projection, we have simulated the linear covariant gauge at $\beta = 2.2$ and
$\beta = 2.3$ for the lattice volumes $V = 8^4, 16^4$ and $24^4$, considering
several values of the gauge parameter $\xi$ in the SU(2) case (see Table
\ref{table-num}). We observe that the quantity $D_l(p^2) p^2/\xi$, which
should be equal to 1, has a value of $0.999(2)$ when averaged over all data
$D_l(p^2)$ produced. From the results shown in Figures \ref{fig:Dt16} and \ref{fig:Dt}
one clearly sees that, as in Landau gauge, the propagator is more infrared
suppressed when the lattice volume increases. At the same time, for a
fixed volume $V$, the propagator is also more infrared suppressed when the
gauge parameter $\xi$ increases. The latter result is in agreement with
Ref.\ \cite{Giusti:2000yc}.

\begin{table}
\vskip -2mm
\caption{Values of the lattice coupling $\beta$, the lattice volume $V$ and the gauge
         parameter $\xi$ used in our simulations. We also report, in each case,
         the total number of configurations considered.
\label{table-num}}
\vskip 2mm \begin{center}
\begin{tabular}{cccc|cccc}
\hline
 $\beta$ & $V$ & $\xi$ & number of conf.\ & $\beta$ & $V$ & $\xi$ & number of conf.\ \\
 \hline
 2.2 &  $8^4$ & 0.0  & 500 & 2.3 &  $8^4$ & 0.0  & 500 \\
 2.2 &  $8^4$ & 0.01 & 500 & 2.3 &  $8^4$ & 0.01 & 500 \\
 2.2 &  $8^4$ & 0.05 & 500 & 2.3 &  $8^4$ & 0.05 & 500 \\
 2.2 &  $8^4$ & 0.1  & 500 & 2.3 &  $8^4$ & 0.1  & 500 \\
 2.2 &  $8^4$ & 0.2  & 500 & 2.3 &  $8^4$ & 0.3  & 500 \\
 2.2 &  $8^4$ & 0.3  & 500 & 2.3 &  $8^4$ & 0.5  & 500 \\
 2.2 &  $8^4$ & 0.4  & 500 & 2.3 &  $8^4$ & 0.6  & 500 \\
 2.2 &  $8^4$ & 0.5  & 542 & 2.3 &  $8^4$ & 0.7  & 543 \\
 \hline
 2.2 & $16^4$ & 0.0  & 400 & 2.3 & $16^4$ & 0.0  & 400 \\
 2.2 & $16^4$ & 0.01 & 400 & 2.3 & $16^4$ & 0.01 & 400 \\
 2.2 & $16^4$ & 0.05 & 400 & 2.3 & $16^4$ & 0.05 & 400 \\
 2.2 & $16^4$ & 0.1  & 400 & 2.3 & $16^4$ & 0.1  & 400 \\
     &        &      &     & 2.3 & $16^4$ & 0.2  & 336 \\
 \hline
 2.2 & $24^4$ & 0.0  & 158 & 2.3 & $24^4$ & 0.0  & 200 \\
 2.2 & $24^4$ & 0.01 & 200 & 2.3 & $24^4$ & 0.01 & 200 \\
 2.2 & $24^4$ & 0.05 & 160 & 2.3 & $24^4$ & 0.05 & 310 \\
     &        &      &     & 2.3 & $24^4$ & 0.07 &  59 \\
 \hline
\end{tabular}
\end{center}
\vskip 1mm
\end{table}

Let us recall that, on the lattice, the finite size of the system corresponds
to an infrared cutoff $\sim 2 \pi/L$, where $L$ is the lattice size. In the
four-dimensional case for $\beta = 2.2$, in Landau gauge, one needs to use a
lattice volume $V$ of about $64^4$ in order to obtain infinite-volume-limit
results \cite{Cucchieri:2007rg,Cucchieri:2008fc}. The data shown in Figures
\ref{fig:Dt16} and \ref{fig:Dt} seem to indicate that similar lattice volumes
are also needed for the linear covariant gauge. On the other hand,
considering the data reported in Section \ref{sec:dis}, the extrapolation
to infinite volume for a given $\beta$ and a fixed value of $\xi$ seems
harder than in Landau gauge. Indeed, as $V \to \infty$,
the number of sites characterized by a large value for the function
$\Lambda^b(x)$ increases, making the convergence of the gauge-fixing method
more difficult.

\section{Conclusions}
\label{sec:concl}

We have discussed a new lattice implementation of the linear covariant gauge.
The gauge fixing is done using a minimizing functional ${\cal E}_{LCG}\{U^{g}, g\}$,
which is a simple generalization of the Landau-gauge functional
${\cal E}_{LG}\{U^{g}\}$. Tests done for the SU(2) case in four space-time dimensions
show that this approach solves most problems encountered in earlier implementations
and ensures a good quality for the gauge fixing with a ratio $D_l(p^2) p^2/\xi \approx 1$
for all cases considered. We have also presented preliminary results for the transverse
gluon propagator $D_t(p^2)$.

As discussed in Sections \ref{sec:dis} and \ref{sec:results}, the only open problem
is the convergence of the gauge-fixing algorithm at large lattice volumes
when the gauge parameter $\xi$ is also large. However, as mentioned above, this
problem is probably less severe for the SU(3) group compared to the SU(2) case. 
We are currently simulating other values of $\beta$ and $\xi$ in the 4d SU(2) case
and considering simulations also of the SU(3) group and of the 3d case \cite{preparation}.


\section*{Acknowledgement}

A.C. and T.M. acknowledge partial support from CNPq.
Support from FAPESP (under grant \# 2009/50180-0) is also acknowledged.
The work of T.M. was supported also by the Alexander von Humboldt Foundation.
E.M.S.S. acknowledges support from CAPES.



\end{document}